\documentclass[prl,aps,twocolumn]{revtex4-1}
\usepackage{amssymb}
\usepackage{amsmath}
\usepackage{graphicx}
\usepackage{color}
\date{\today}

\begin{document}

\title{Boundary-Driven Anomalous Spirals in Oscillatory Media}

\author{David A. Kessler}
\affiliation{Department of Physics, Bar-Ilan University, Ramat-Gan IL52900, Israel}

\author{Herbert Levine }
\affiliation{Department of Bioengineering, Center for Theoretical Biological Physics, Rice University, Houston, TX 77005, U.S.A.}

\begin{abstract}
We study a heretofore ignored class of spiral patterns for oscillatory media as characterized by the complex Landau-Ginzburg model. These spirals emerge from modulating the growth rate as a function of $r$, thereby turning off the instability. These spirals are uniquely determined by matching to those outer conditions, lifting a degeneracy in the set of steady-state solutions of the original equations. Unlike the well-studied spiral which acts a wave source, has a simple core structure and is insensitive to the details of the boundary on which no-flux conditions are imposed, these new spirals are wave sinks, have non-monotonic wavefront curvature near the core, and can be patterned by the form of the spatial boundary. We predict that these anomalous spirals could be produced in nonlinear optics experiments via spatially modulating the gain of the medium.
\end{abstract}

\maketitle

Oscillatory media, wherein the equilibrium state becomes unstable to a spatially uniform oscillatory mode, i.e. undergoes a Hopf bifurcation, constitute an important class of non-equilibrium systems. Examples of such systems range from chemical dynamics \cite{epstein} to lasers \cite{coullet,bazhenov} to living matter \cite{goldbeter}. 
Close to the onset of the instability, all of these systems can be described by a universal model, the complex Ginsburg-Landau equation (CGLE)~\cite{CrossHoh}. Assuming the bifurcation is supercritical, the CGLE takes the form
\begin{equation}
\dot{U} = (1+ib)\nabla^2 U  + U - (1+ic) |U|^2 U
\label{CGLE}
\end{equation}
For definiteness, we will focus herein on the case $b=0$, though the conclusions we will reach apply equally well to the more general case. This equation has been the subject of many studies (for a review see \cite{Aranson}) as it exhibits a variety of interesting non-linear pattern-forming properties.  One basic aspect of this phenomenology is the spiral wave pattern, uniformly rotating at a constant angular frequency $\omega$.  The spiral solution takes the form, in two-dimensional polar coordinates $r$, $\phi$,
\begin{equation}
U(r,\phi) = \rho(r) e^{-i\omega t + im\phi + i\psi(r)}
\label{spiral}
\end{equation}
The integer $m$ denotes the topological charge of the spiral, i.e. by what multiple of $2\pi$ the phase increases as we traverse a closed path circling the origin in counterclockwise fashion. At large $r$ the spiral reduces to the plane wave solution 
\begin{equation}
\psi'(r) \to q, \qquad \rho \to \sqrt{1-q^2}
\end{equation} This implies the dispersion relation
\begin{equation}
\omega = c(1-  q^2)\label{disperse}
\end{equation}

This spiral wave pattern was studied in depth by Hagan in 1982~\cite{Hagan}.  Hagan was able to construct an analytic approximation for the spiral in the limit of small $c$.  Interestingly, the Hagan solution contained a free parameter, denoted by him as $\alpha$.  The $\alpha=0$ solution was qualitatively different from the finite $\alpha$ solutions.  In particular, the $\alpha=0$ solution was what has come to be called an ``antispiral" ~\cite{Vanag,Brusch} whose sense of rotation is the same as the direction that the spiral wraps outward around the origin.  The solutions for $\alpha\ne 0$, however, were ``spirals", where the sense of rotation is the opposite of the wrapping direction. Moreover, whereas the $\alpha=0$ antispiral has a phase which increased monotonically (in absolute value) with increasing radius $r$, the $\alpha\ne 0$ spirals, had a non-monotonic dependence of phase on $r$. The $\alpha=0$ antispiral has been seen many times in simulations of systems with both no-flux and periodic boundary conditions, and agrees quantitatively with Hagan's theory.  The $\alpha\ne 0$ spiral has hardly been discussed, and to the best of our knowledge has not previously been seen in simulation; the prevailing assumption seems to be that these solution are linearly unstable ~\cite{krinskii} In this paper, we show how, with appropriate boundary conditions, an $\alpha\ne 0$ spiral can be generated.  Moreover, we elucidate the mechanism whereby a particular $\alpha$ (or equivalently $\omega$ or asymptotic wavenumber $q$) is selected out of the continuum of possible steady-state solutions.  This mechanism, namely boundary driven selection, implies that the resulting spiral has a number of anomalous features which we explore.

If we plug the spiral ansatz, Eq. (\ref{spiral}), into the CGLE, we find that the functions $\rho(r)$, $\psi'(r)$ must satisfy the equations
\begin{align}
0&=\rho'' + \frac{\rho'}{r} - \frac{m^2\rho}{r^2} - \rho\psi'^2 + \rho - \rho^3\nonumber\\
-\omega\rho &= \rho\psi'' + \frac{\rho\psi'}{r} + 2\rho'\psi' - c\rho^3
\end{align}
The signature characteristic of the Hagan $\alpha=0$ spiral is that $cq<0$.  Then, linearizing about the large $r$ solution reveals two modes that grow exponentially with $r$ and one decaying mode.  A similar analysis at small $r$ indicates that there is one free parameter characterizing the small $r$ behavior, namely $\rho'(0)$. The two growing modes must not be present in the solution and so we have to fix not only $\rho'(0)$ but also $\omega$ to obtain a solution for the entire infinite range of $r$. This counting argument is the origin of the unique $\omega$ associated with the Hagan $\alpha=0$ solution. For this solution, taking, for example, $c>0$, we have $q<0$ and the spiral wraps around the origin in a counterclockwise fashion.  We have by Eq. (\ref{disperse}), $\omega>0$
so the spiral also rotates counterclockwise, and the solution is an antispiral. At large $r$ we have waves with negative, i.e. incoming, phase velocity.  However, since $d\omega/dq<0$, these waves have  a positive (i.e. outgoing) group velocity, and the antispiral is a wave source. Hagan found that the $m=\pm 1$ antispirals were stable, while the higher $m$ solutions were not, so we restrict our attention in the following to this case.

\begin{figure}
 \begin{center}
 \includegraphics[width=0.45\textwidth]{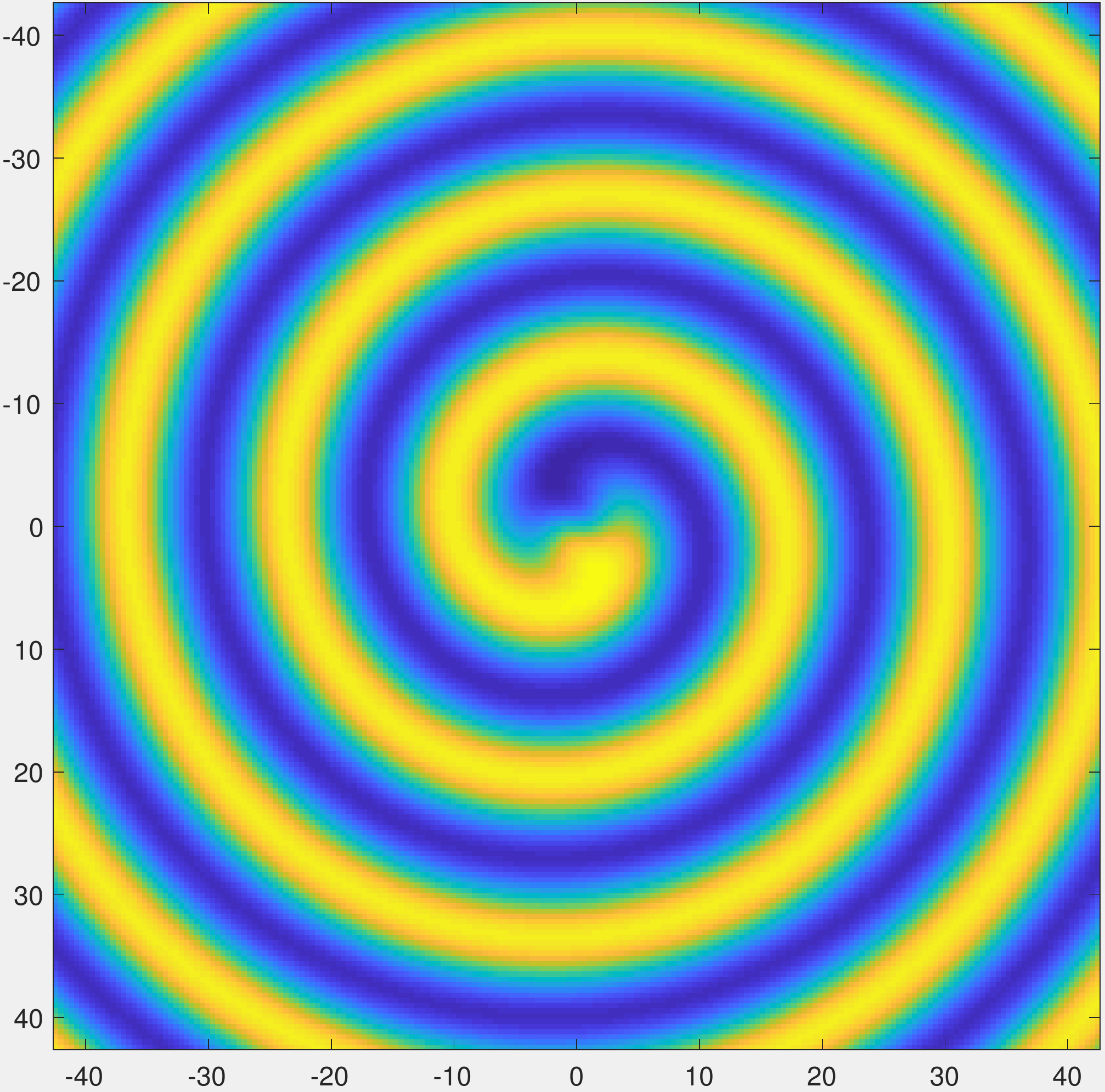}\\
 \includegraphics[width=0.45\textwidth]{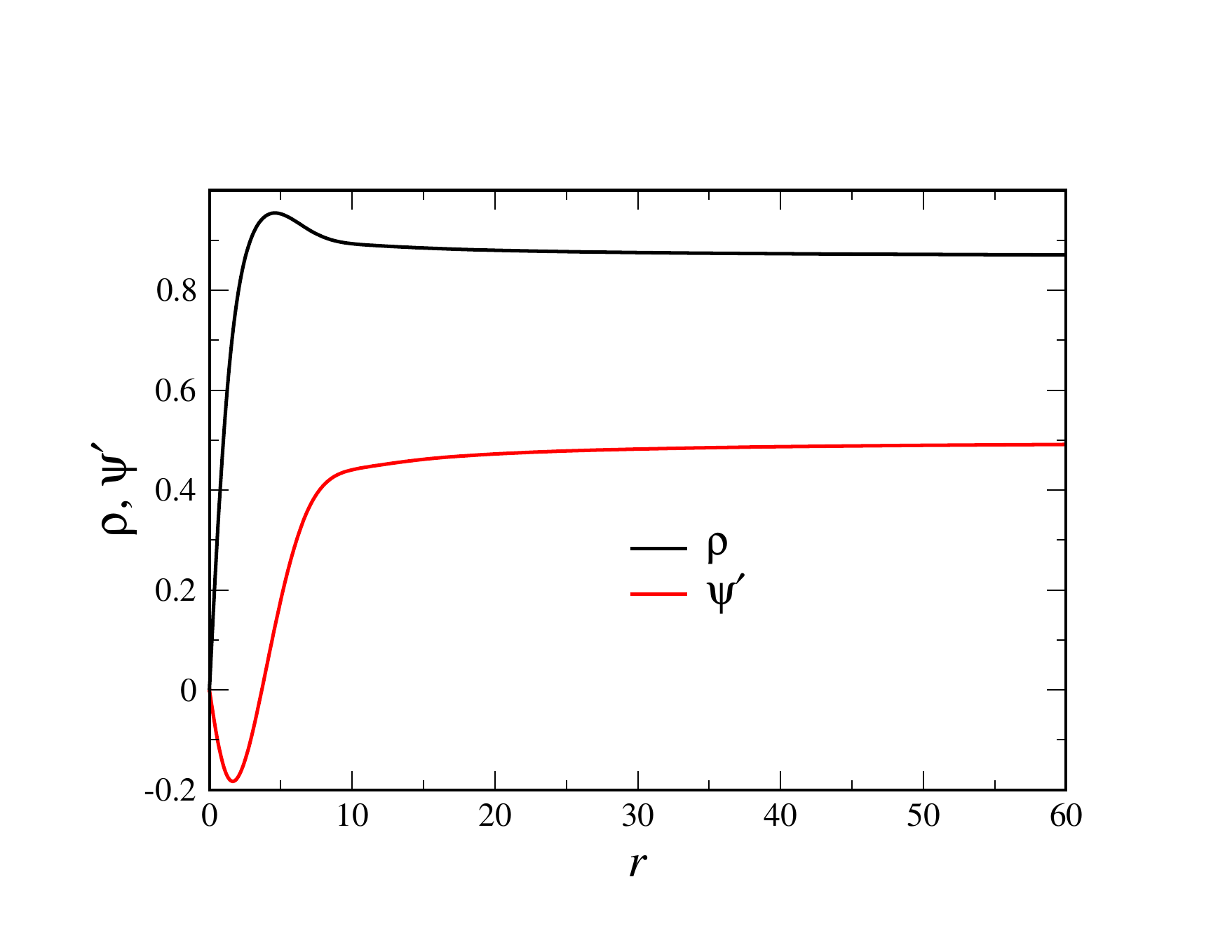}
 \end{center}
 \caption{The single-armed Hagan $\alpha\ne 0$ spiral solution with $c=1$, $\omega=0.75$. Left: $\Re(U)$. Right: $\rho(r)$, $\psi(r)$. Notice should be taken off the two maxima of curvature in the core region and the associated non-monotonic behavior of $\psi(r)$.}
 \label{fig1}
 \end{figure}

The $\alpha\ne 0 $ solutions have $cq>0$.  The analysis at large $r$ now indicates two decaying modes, and so a continuum of solutions with different $\omega$s exist.  It is a simple matter to generate these solutions via a collocation method. One fixes $\omega$ and takes as an initial guess a $\rho$ which increases linearly at the origin and saturates for large $r$, and a $\psi'$ which has the correct behavior at small and large $r$. The collocation method then discretizes $r$ in the range from some $r_0\ll 1$ to $r_f \gg 1$, and approximates the differential equation at the interior points $r_i$ as a set of nonlinear equations in the unknowns $\rho(r_i)$, $\psi'(r_i)$.  These equations are supplemented by a set of boundary conditions, namely that $\rho(r_0) = r_0 
\rho'(r_0)$, $\psi'(r_0)= - \omega r_0/4$ and $\rho'(r_f) = 0$.  The first two of these guarantee the smoothness of the solution at small $r$ and the last the well-behavedness at large $r$, ensuring the absence of the three bad modes. This set of nonlinear equations is then fed to a nonlinear solver such as Newton's method. In practice, we employ the Matlab routine {\em{bvp4c}}, and find convergence to the desired solution.  
 A portrait of one such solution is shown in Fig. \ref{fig1}. We see that the solution has an interesting core structure, with two local maxima of the curvature, as opposed to the unique maximum curvature tip of the $\alpha=0$ antispiral.  This is related to the non-monotonicity of $\psi$, which for $c>0$, starts out negative and then turns up. The solution is a spiral, with a clockwise wrapping for $c>0$, opposite to the sense of rotation.  It has outgoing phase velocity, and so incoming group velocity, waves at large $r$. Taking one of these solutions as an initial condition with periodic or zero-flux boundary conditions gives rise to a dynamics which after a transient, relaxes back to the $\alpha=0$ solution.  Thus raises the question of whether it is possible to stabilize the $\alpha\ne 0$ solution.  And, if so, is there a unique selected $\omega$, or does $\omega$ depend on the initial conditions?

\begin{figure}
 \begin{center}
 \includegraphics[width=0.45\textwidth]{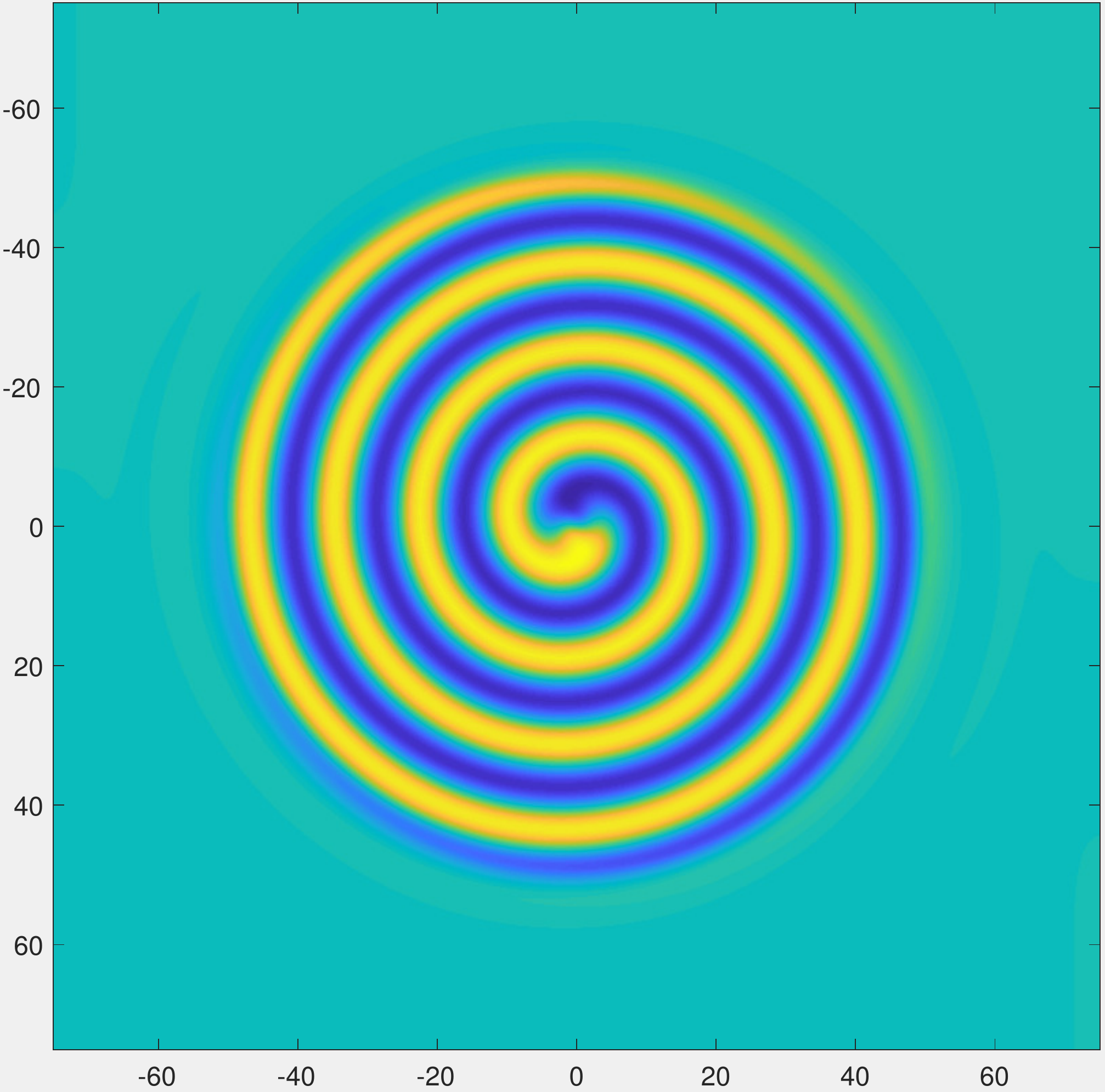}
 \end{center}
 \caption{The long time pattern of $\Re(U)$ using a $\tanh$ cutoff with $r_c=50$, $c=1$.  The initial condition was a Hagan $\alpha=0$ antispiral. The spiral rotates counterclockwise.}
 \label{fig2}
 \end{figure}

The answer to both these questions lies in changing the counting argument, i.e., modifying the linearized equation at large $r$.  One convenient way to do this is to ``turn off" the linear instability past some critical radius.  For example, one can multiply the $U$ term in the CGLE, Eq. (\ref{CGLE}), by a factor
$\gamma(r) = (1 + \tanh(r_c - r))/2$. This is analogous to the idea of adding a parameter ramp connecting the actual system to one which is below the stability threshold; we will return to this connection later. A simulation of the CGLE with this modification yields the spiral in Fig. \ref{fig2}.  This looks very similar to the $\alpha\ne 0$ spiral as found by the collocation method, and shares all its major features.  It is a spiral, has outgoing phase waves, and has two maxima of the curvature.  It is interesting to note that at large enough $r$, the field stops rotating, presumably due to the interaction with the outer boundary at $|x|=|y|=L/2$. This does not however prevent the pattern for $r<r_c$ from rotating essentially uniformly.  Also, most crucially, the frequency of the spiral is independent of initial conditions, as well as the box size $L$. The same general features hold independent of the width of the transition to the cutoff region, which can be made as sharp as desired, including using a step function $\gamma(r)=\theta(r_c-r)$. In fact, simply clamping the field to $U=0$ on the boundary is sufficient to induce an $\alpha\ne 0$ spiral, with frequency selection.  The precise value of the selected frequency of course depends on the details of the cutoff.

In order to get an analytic handle on the selection problem, one can multiply both the linear and cubic terms by $\gamma (r)$, with the same qualitative results.  For the purpose of analysis, it is simplest to consider the step function form of $
\gamma$ discussed above. In this case, for $r>r_c$, we can solve exactly for the rotating solution of the resulting diffusion equation
\begin{equation}
\dot{U} =-i\omega U = \nabla^2 U
\end{equation}
namely,
\begin{equation}
U(\rho,\phi) = AK_{|m|}\left((1-i)\sqrt{\frac{\omega}{2}}r\right)e^{im\phi-i\omega t}
\end{equation}
This gives us two boundary conditions on $U$ at $r_c$, one specifying $\rho'(r_c)/\rho)$ and the other $\psi'(r_c)$. We then need to adjust $\rho'(0)$ and $\omega$ to construct a solution, giving us a selection criterion for $\omega$. It is straightforward to implement this by taking as an initial guess an $\alpha\ne 0$ spiral without any cutoff, and using collocation to determine the solution along with $\omega$.  The $\omega$ calculated in this fashion matches exactly the rotation frequency of the full simulation of the time-dependent system.  Given the solution for the step function $\gamma(r)$, it is straightforward to continually modify the form of $\gamma$ to obtain the selected $\omega$ for more general forms of $\gamma$, such as the $\tanh$ function described earlier.

\begin{figure}
 \begin{center}
 \includegraphics[width=0.45\textwidth]{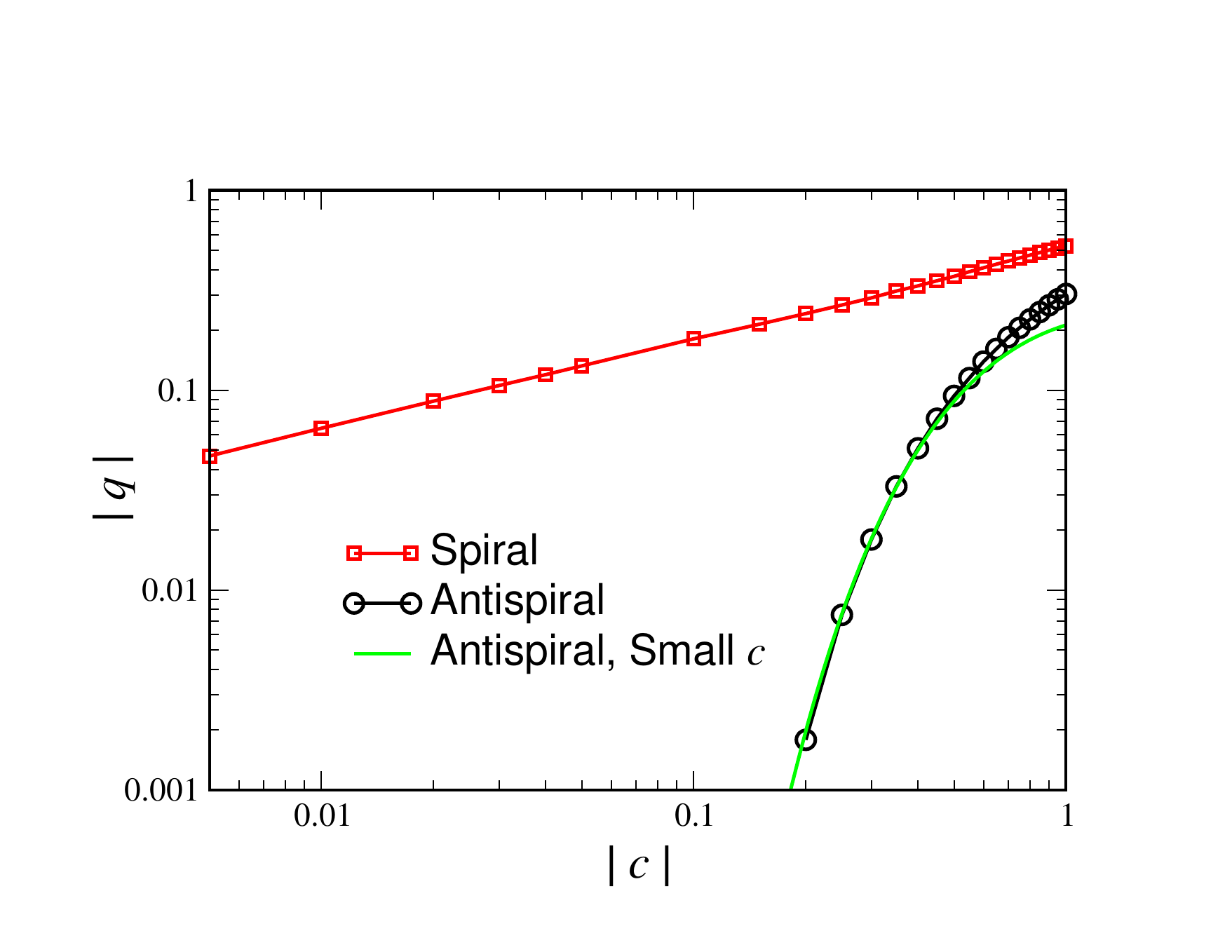}
 \end{center}
 \caption{The selected asymptotic wave number vs. $c$ in the $r_c\to \infty$ limit with a step-function cutoff on the linear and cubic reaction terms. For comparison, the results for Hagan $\alpha=0$ antispiral are also shown, along with Hagan's small $c$ analytic approximation}
 \label{fig3}
 \end{figure}

One can now ask what is the dependence of the selected $\omega$ on the cutoff radius $r_c$. We find that $\omega$ increases mildly with $r_c$ and approaches a constant as $r_c\to \infty$.  This points out the singular nature of the selection mechanism, as a cutoff at infinite $r_c$ is not equivalent to no cutoff at all.  In the large $r_c$ limit, the selection problem simplifies greatly, since the spiral follows the non-cutoff solution until a short distance before $r_c$ and then deviates, following the one growing mode of the large-$r$ analysis, and matches on to the cutoff boundary conditions, which in the large $r_c$ limit are simply
$\rho'(r_c)/\rho(r_c) = -\psi'(r_)=-\sqrt{\omega/2}$.  Since $r$ is large in this entire region, the spiral equations reduce to
\begin{align}
0&=\rho'' - \rho\psi'^2 + \rho - \rho^3\nonumber\\
-\omega\rho &= \rho\psi'' + 2\rho'\psi' - c\rho^3
\label{infR}
\end{align}
with the boundary conditions $\rho \to \sqrt{1-q^2}$, $\psi'\to q$ as $r\to -\infty$, and the above conditions at $r=r_c$.  This can be easily solved by standard shooting methods
upon exploiting the translation symmetry in $r$. We
integrate from some large negative $r-r_c$  until the condition on $\rho'/\rho=-
\sqrt{\omega/2}$ is satisfied, calling this point $r_c$, and varying $\omega$ until the condition on $\psi'(r_c)$ is also satisfied.  Doing this, we calculate $\omega(c)$, presenting the results for $q=\sqrt{1-\omega/c}$ of the single-armed spiral ($m=1$) in Fig. \ref{fig3}, together with the $|q|$ for the selected $\omega$ for the $\alpha=0$, $m=1$ antispiral for comparison.  We see that the $q$ of the $\alpha\ne 0$ is larger than that of the antispiral, especially for small $c$, where $|q|$ of the antispiral vanishes as
$\exp(-\pi/(2c))$ for small $c$.  For the spiral, on the other hand, it appears that $q$ is proportional to $\sqrt{c}$ as $c\to 0$.

It should be noted that the selection equation Eq. (\ref{infR}) is precisely the equation for the selection of the frequency (and hence wavenumber) for a one-dimensional wave pattern where the linear and cubic reaction terms have been turned off beyond some maximal $|x|$. As already mentioned, the possibility of inducing wavelength selection in a one dimensional system via modulating the system parameters in space so that the system is subcritical near the boundary is a well-studied technique~\cite{CrossHoh}.  What has been shown here is an extension of this technique to the problem of spirals in two dimensions, where here it induces an altogether qualitatively different pattern.

\begin{figure*}
 \begin{center}
 \includegraphics[width=0.45\textwidth]{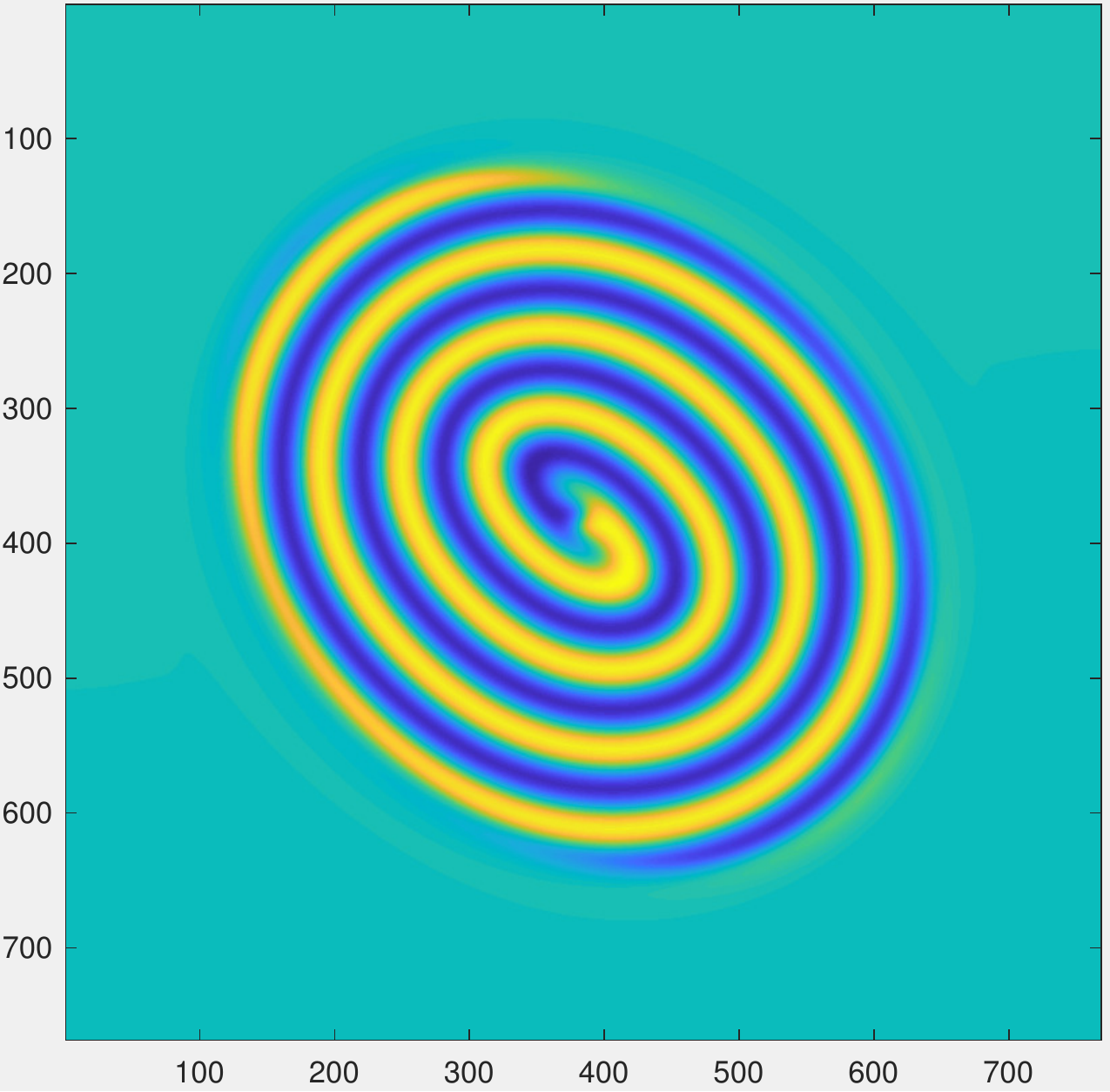}
 \includegraphics[width=0.45\textwidth]{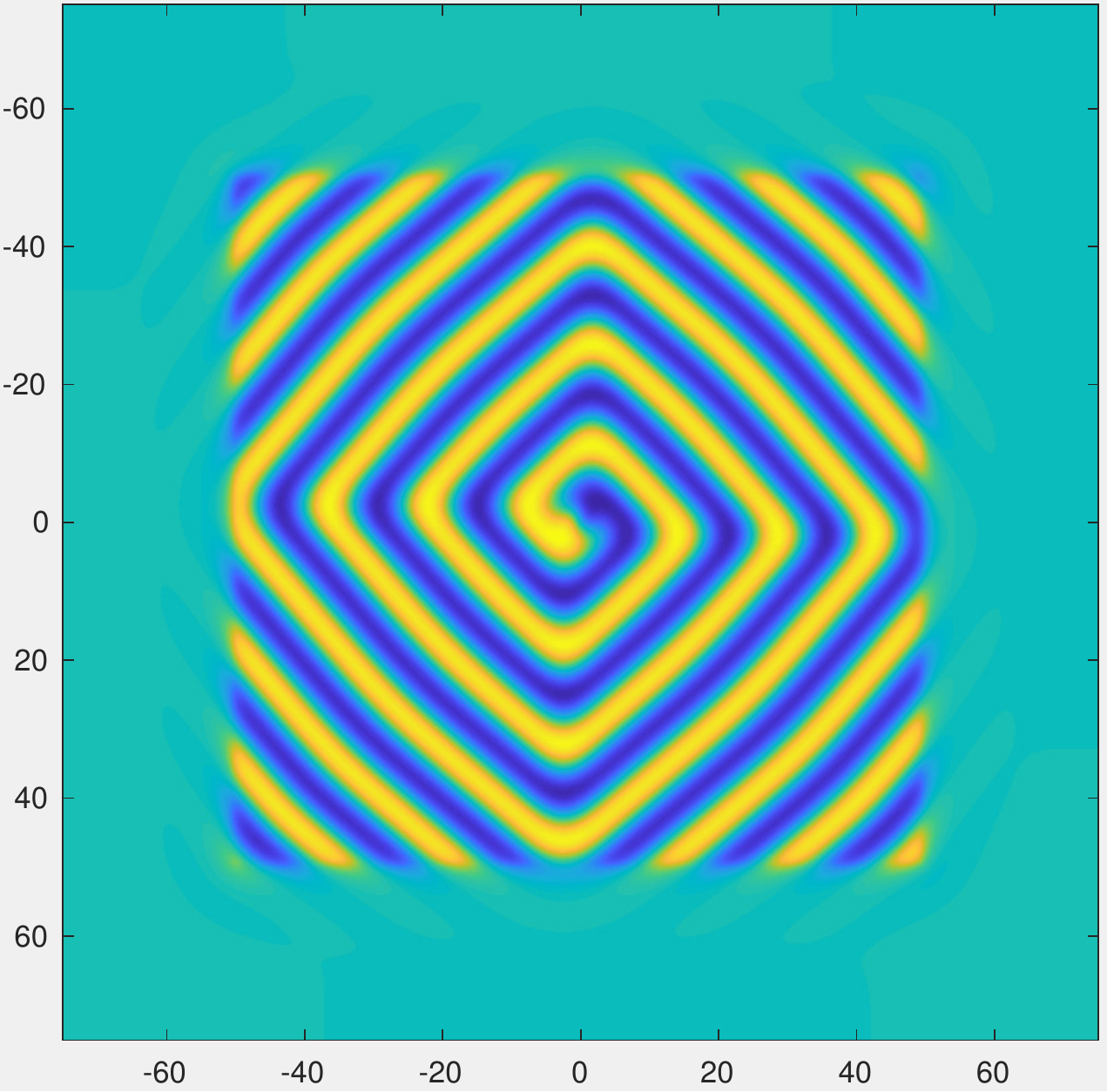}
 \end{center}
 \caption{The long-time pattern of $\Re(U)$ for simulations with the tanh cutoffs on the linear and cubic reaction, and different forms of $r_c(\phi)$, with $c=1$. 
 The initial condition was the Hagan $\alpha=0$ spiral. Left: $r_c(\phi)$ traces out a $45^\circ$ tilted ellipse with $(r_++r_-)/2=50$ and eccentricity $0.4$. Right: $r_c(\phi)$ traces out the square with edge length 100 centered on the origin. }
 \label{fig4}
 \end{figure*}

Given that the spatial modulation of the growth rate fixes the pattern, one may ask what is the effect of a parameter modulation which breaks the radial symmetry.  For example, one may consider a case where $r_c$ is a given function of $\phi$, say a tilted elliptical form, $r_c(\phi)=(r_+r_-)/\sqrt{r_-^2\cos^2(\phi -\pi/4)+ r_+^2\sin^2(\phi-\pi/4)}$, where $r_\pm = r_c^* \pm \delta$.  A post-transient snapshot of the resulting pattern is show in the left-hand panel of Fig. \ref{fig4}.  We see that the entire spiral has been distorted into an elliptical shape, due to the influence of the cutoff, but the spiral nevertheless rotates in a periodic fashion.  An even more extreme example is shown in the right-hand panel of this figure.  Here, we chose $r_c(\phi)$ to lie on the circumference of a square of size $r_c^*$ centered at the origin.  Now the spiral adopts a diamond type shape.

Lastly, we turn to the fate of the cutoff system with random initial conditions, say with Gaussian distributed amplitude and random phase. For either periodic or no-flux boundary conditions, the system freezes into a state with many tiny $m=\pm 1$ antispiral pairs.  In the presence of a cutoff, however, the system evolves to a single spiral at the center, due to the forcing of the pattern from the boundaries.

Thus, we have seen that by turning off the instability of a two-dimensional oscillatory medium outside of some central region, we can stabilize a pattern, namely the Hagan $\alpha\ne0$ spiral, that is otherwise not seen in simulation. This form of boundary control is much richer than is typical in one dimensional systems, where the boundary simply induces wavelength selection,~\cite{CrossHoh} but does not qualitatively change the pattern. Furthermore, by varying the details of this parameter modulation, we can exercise some degree of control over the selected frequency.  Varying the shape of the cutoff region affects the entire interior pattern, generating novel patterns hitherto unimaginable. It also imposes order on the system, causing an initial random configuration to produce an ordered final state.  While various other modulations of parameters in the CGLE have been studied previously ~\cite{Li2008,Li2013,ott}, the reversion to the bifurcation point seems to have been overlooked. Gil, et al.~\cite{Gil} in fact introduced a radial modulation of all the parameters, and in particular the outer region was subcritical, but they did not recognize the connection of the resulting pattern to the Hagan $\alpha\ne 0$.  It is interesting to contemplate what other novel patterns can be induced via boundary modification. It is almost certainly the case that these results are not unique to the CGLE and would hold for more detailed models of specific oscillatory systems. Finally, it would be most interesting to try to implement our protocol experimentally.  One possibility is creating a nonlinear optical system in which the gain is modulated top go below unity beyond a critical radius.

\acknowledgments{The research of DAK is supported by the US-Israel Binational Science Foundation under grant 2015619.  The research of HL is supported by the National Science Foundation Center for Theoretical Biological Physics (Grant NSF PHY-1427654). We also gratefully acknowledge the hospitality of the Aspen Center for Physics, where this work was started.}
\bibliography{BoundarySpiral}

\begin{thebibliography}{14}%
\makeatletter
\providecommand \@ifxundefined [1]{%
 \@ifx{#1\undefined}
}%
\providecommand \@ifnum [1]{%
 \ifnum #1\expandafter \@firstoftwo
 \else \expandafter \@secondoftwo
 \fi
}%
\providecommand \@ifx [1]{%
 \ifx #1\expandafter \@firstoftwo
 \else \expandafter \@secondoftwo
 \fi
}%
\providecommand \natexlab [1]{#1}%
\providecommand \enquote  [1]{``#1''}%
\providecommand \bibnamefont  [1]{#1}%
\providecommand \bibfnamefont [1]{#1}%
\providecommand \citenamefont [1]{#1}%
\providecommand \href@noop [0]{\@secondoftwo}%
\providecommand \href [0]{\begingroup \@sanitize@url \@href}%
\providecommand \@href[1]{\@@startlink{#1}\@@href}%
\providecommand \@@href[1]{\endgroup#1\@@endlink}%
\providecommand \@sanitize@url [0]{\catcode `\\12\catcode `\$12\catcode
  `\&12\catcode `\#12\catcode `\^12\catcode `\_12\catcode `\%12\relax}%
\providecommand \@@startlink[1]{}%
\providecommand \@@endlink[0]{}%
\providecommand \url  [0]{\begingroup\@sanitize@url \@url }%
\providecommand \@url [1]{\endgroup\@href {#1}{\urlprefix }}%
\providecommand \urlprefix  [0]{URL }%
\providecommand \Eprint [0]{\href }%
\providecommand \doibase [0]{http://dx.doi.org/}%
\providecommand \selectlanguage [0]{\@gobble}%
\providecommand \bibinfo  [0]{\@secondoftwo}%
\providecommand \bibfield  [0]{\@secondoftwo}%
\providecommand \translation [1]{[#1]}%
\providecommand \BibitemOpen [0]{}%
\providecommand \bibitemStop [0]{}%
\providecommand \bibitemNoStop [0]{.\EOS\space}%
\providecommand \EOS [0]{\spacefactor3000\relax}%
\providecommand \BibitemShut  [1]{\csname bibitem#1\endcsname}%
\let\auto@bib@innerbib\@empty
\bibitem [{\citenamefont {Epstein}\ and\ \citenamefont
  {Pojman}(1998)}]{epstein}%
  \BibitemOpen
  \bibfield  {author} {\bibinfo {author} {\bibfnamefont {I.~R.}\ \bibnamefont
  {Epstein}}\ and\ \bibinfo {author} {\bibfnamefont {J.~A.}\ \bibnamefont
  {Pojman}},\ }\href@noop {} {\emph {\bibinfo {title} {An introduction to
  nonlinear chemical dynamics: oscillations, waves, patterns, and chaos}}}\
  (\bibinfo  {publisher} {Oxford University Press},\ \bibinfo {address}
  {Oxford},\ \bibinfo {year} {1998})\BibitemShut {NoStop}%
\bibitem [{\citenamefont {Coullet}\ \emph {et~al.}(1989)\citenamefont
  {Coullet}, \citenamefont {Gil},\ and\ \citenamefont {Rocca}}]{coullet}%
  \BibitemOpen
  \bibfield  {author} {\bibinfo {author} {\bibfnamefont {P.}~\bibnamefont
  {Coullet}}, \bibinfo {author} {\bibfnamefont {L.}~\bibnamefont {Gil}}, \ and\
  \bibinfo {author} {\bibfnamefont {F.}~\bibnamefont {Rocca}},\ }\href@noop {}
  {\bibfield  {journal} {\bibinfo  {journal} {Optics Communications}\ }\textbf
  {\bibinfo {volume} {73}},\ \bibinfo {pages} {403} (\bibinfo {year}
  {1989})}\BibitemShut {NoStop}%
\bibitem [{\citenamefont {Bazhenov}\ \emph {et~al.}(1990)\citenamefont
  {Bazhenov}, \citenamefont {Vasnetsov},\ and\ \citenamefont
  {Soskin}}]{bazhenov}%
  \BibitemOpen
  \bibfield  {author} {\bibinfo {author} {\bibfnamefont {V.~Y.}\ \bibnamefont
  {Bazhenov}}, \bibinfo {author} {\bibfnamefont {M.}~\bibnamefont {Vasnetsov}},
  \ and\ \bibinfo {author} {\bibfnamefont {M.}~\bibnamefont {Soskin}},\
  }\href@noop {} {\bibfield  {journal} {\bibinfo  {journal} {{JETP Lett.}}\
  }\textbf {\bibinfo {volume} {52}},\ \bibinfo {pages} {429} (\bibinfo {year}
  {1990})}\BibitemShut {NoStop}%
\bibitem [{\citenamefont {Goldbeter}(1997)}]{goldbeter}%
  \BibitemOpen
  \bibfield  {author} {\bibinfo {author} {\bibfnamefont {A.}~\bibnamefont
  {Goldbeter}},\ }\href@noop {} {\emph {\bibinfo {title} {Biochemical
  oscillations and cellular rhythms: The molecular bases of periodic and
  chaotic behaviour}}}\ (\bibinfo  {publisher} {Cambridge University Press},\
  \bibinfo {address} {Cambridge},\ \bibinfo {year} {1997})\BibitemShut
  {NoStop}%
\bibitem [{\citenamefont {Cross}\ and\ \citenamefont
  {Hohenberg}(1993)}]{CrossHoh}%
  \BibitemOpen
  \bibfield  {author} {\bibinfo {author} {\bibfnamefont {M.~C.}\ \bibnamefont
  {Cross}}\ and\ \bibinfo {author} {\bibfnamefont {P.~C.}\ \bibnamefont
  {Hohenberg}},\ }\href@noop {} {\bibfield  {journal} {\bibinfo  {journal}
  {Rev. Mod. Phys.}\ }\textbf {\bibinfo {volume} {65}},\ \bibinfo {pages} {851}
  (\bibinfo {year} {1993})}\BibitemShut {NoStop}%
\bibitem [{\citenamefont {Aranson}\ and\ \citenamefont
  {Kramer}(2002)}]{Aranson}%
  \BibitemOpen
  \bibfield  {author} {\bibinfo {author} {\bibfnamefont {I.~S.}\ \bibnamefont
  {Aranson}}\ and\ \bibinfo {author} {\bibfnamefont {L.}~\bibnamefont
  {Kramer}},\ }\href@noop {} {\bibfield  {journal} {\bibinfo  {journal} {Rev.
  Mod. Phys.}\ }\textbf {\bibinfo {volume} {74}},\ \bibinfo {pages} {99}
  (\bibinfo {year} {2002})}\BibitemShut {NoStop}%
\bibitem [{\citenamefont {Hagan}(1982)}]{Hagan}%
  \BibitemOpen
  \bibfield  {author} {\bibinfo {author} {\bibfnamefont {P.~S.}\ \bibnamefont
  {Hagan}},\ }\href@noop {} {\bibfield  {journal} {\bibinfo  {journal} {SIAM J.
  Appl. Math.}\ }\textbf {\bibinfo {volume} {42}},\ \bibinfo {pages} {762}
  (\bibinfo {year} {1982})}\BibitemShut {NoStop}%
\bibitem [{\citenamefont {Vanag}\ and\ \citenamefont {Epstein}(2001)}]{Vanag}%
  \BibitemOpen
  \bibfield  {author} {\bibinfo {author} {\bibfnamefont {V.~K.}\ \bibnamefont
  {Vanag}}\ and\ \bibinfo {author} {\bibfnamefont {I.~R.}\ \bibnamefont
  {Epstein}},\ }\href@noop {} {\bibfield  {journal} {\bibinfo  {journal}
  {Science}\ }\textbf {\bibinfo {volume} {294}},\ \bibinfo {pages} {835}
  (\bibinfo {year} {2001})}\BibitemShut {NoStop}%
\bibitem [{\citenamefont {Brusch}\ \emph {et~al.}(2003)\citenamefont {Brusch},
  \citenamefont {Nicola},\ and\ \citenamefont {B{\"a}r}}]{Brusch}%
  \BibitemOpen
  \bibfield  {author} {\bibinfo {author} {\bibfnamefont {L.}~\bibnamefont
  {Brusch}}, \bibinfo {author} {\bibfnamefont {E.~M.}\ \bibnamefont {Nicola}},
  \ and\ \bibinfo {author} {\bibfnamefont {M.}~\bibnamefont {B{\"a}r}},\
  }\href@noop {} {\bibfield  {journal} {\bibinfo  {journal} {Phys. Rev. Lett.}\
  }\textbf {\bibinfo {volume} {92}},\ \bibinfo {pages} {089801} (\bibinfo
  {year} {2003})}\BibitemShut {NoStop}%
\bibitem [{\citenamefont {Krinskii}\ \emph {et~al.}(1984)\citenamefont
  {Krinskii}, \citenamefont {Mikhailov}, \citenamefont {Panfilov},
  \citenamefont {Ermakova},\ and\ \citenamefont {Tsyganov}}]{krinskii}%
  \BibitemOpen
  \bibfield  {author} {\bibinfo {author} {\bibfnamefont {V.~I.}\ \bibnamefont
  {Krinskii}}, \bibinfo {author} {\bibfnamefont {A.~S.}\ \bibnamefont
  {Mikhailov}}, \bibinfo {author} {\bibfnamefont {A.~V.}\ \bibnamefont
  {Panfilov}}, \bibinfo {author} {\bibfnamefont {E.~A.}\ \bibnamefont
  {Ermakova}}, \ and\ \bibinfo {author} {\bibfnamefont {M.~A.}\ \bibnamefont
  {Tsyganov}},\ }\href@noop {} {\bibfield  {journal} {\bibinfo  {journal}
  {Radiophysics and Quantum Electronics}\ }\textbf {\bibinfo {volume} {27}},\
  \bibinfo {pages} {783} (\bibinfo {year} {1984})}\BibitemShut {NoStop}%
\bibitem [{\citenamefont {Li}\ \emph {et~al.}(2008)\citenamefont {Li},
  \citenamefont {Zhang}, \citenamefont {Ying}, \citenamefont {Chen},\ and\
  \citenamefont {Hu}}]{Li2008}%
  \BibitemOpen
  \bibfield  {author} {\bibinfo {author} {\bibfnamefont {B.-W.}\ \bibnamefont
  {Li}}, \bibinfo {author} {\bibfnamefont {H.}~\bibnamefont {Zhang}}, \bibinfo
  {author} {\bibfnamefont {H.-P.}\ \bibnamefont {Ying}}, \bibinfo {author}
  {\bibfnamefont {W.-Q.}\ \bibnamefont {Chen}}, \ and\ \bibinfo {author}
  {\bibfnamefont {G.}~\bibnamefont {Hu}},\ }\href@noop {} {\bibfield  {journal}
  {\bibinfo  {journal} {Phys. Rev. E}\ }\textbf {\bibinfo {volume} {77}},\
  \bibinfo {pages} {056207} (\bibinfo {year} {2008})}\BibitemShut {NoStop}%
\bibitem [{\citenamefont {Li}\ and\ \citenamefont {Li}(2013)}]{Li2013}%
  \BibitemOpen
  \bibfield  {author} {\bibinfo {author} {\bibfnamefont {T.-C.}\ \bibnamefont
  {Li}}\ and\ \bibinfo {author} {\bibfnamefont {B.-W.}\ \bibnamefont {Li}},\
  }\href@noop {} {\bibfield  {journal} {\bibinfo  {journal} {Chaos}\ }\textbf
  {\bibinfo {volume} {23}},\ \bibinfo {pages} {033130} (\bibinfo {year}
  {2013})}\BibitemShut {NoStop}%
\bibitem [{\citenamefont {Hendrey}\ \emph {et~al.}(2000)\citenamefont
  {Hendrey}, \citenamefont {Ott},\ and\ \citenamefont {Antonsen~Jr}}]{ott}%
  \BibitemOpen
  \bibfield  {author} {\bibinfo {author} {\bibfnamefont {M.}~\bibnamefont
  {Hendrey}}, \bibinfo {author} {\bibfnamefont {E.}~\bibnamefont {Ott}}, \ and\
  \bibinfo {author} {\bibfnamefont {T.~M.}\ \bibnamefont {Antonsen~Jr}},\
  }\href@noop {} {\bibfield  {journal} {\bibinfo  {journal} {Physical Review
  E}\ }\textbf {\bibinfo {volume} {61}},\ \bibinfo {pages} {4943} (\bibinfo
  {year} {2000})}\BibitemShut {NoStop}%
\bibitem [{\citenamefont {Gil}\ \emph {et~al.}(1992)\citenamefont {Gil},
  \citenamefont {Emilsson},\ and\ \citenamefont {Oppo}}]{Gil}%
  \BibitemOpen
  \bibfield  {author} {\bibinfo {author} {\bibfnamefont {L.}~\bibnamefont
  {Gil}}, \bibinfo {author} {\bibfnamefont {K.}~\bibnamefont {Emilsson}}, \
  and\ \bibinfo {author} {\bibfnamefont {G.-L.}\ \bibnamefont {Oppo}},\
  }\href@noop {} {\bibfield  {journal} {\bibinfo  {journal} {Phys. Rev. A}\
  }\textbf {\bibinfo {volume} {45}},\ \bibinfo {pages} {R567} (\bibinfo {year}
  {1992})}\BibitemShut {NoStop}%
\end{thebibliography}%
\end{document}